# Frequency downshifting stair for ultra-intense femtosecond lasers through a plasma-photonics structure

Yunxiao He[1,2], Xiaonan Ning[3], Bo Guo[3], Jianfei Hua[1], Yuqiu Gu[2,4*], and Wei Lu[5,3,1†]

1. Department of Engineering Physics, Tsinghua University, Beijing 100084, China
2. National Key Laboratory of Plasma Physics, Laser Fusion Research Center, China Academy of Engineering Physics, Mianyang, Sichuan 621900, China
3. Beijing Academy of Quantum Information Sciences, Beijing 100193, China
4. Shanghai Institute of Laser Plasma, China Academy of Engineering Physics, Shanghai 201800, China
5. Institute of High Energy Physics, Chinese Academy of Sciences, Beijing 100049, China

Corresponding authors: *yqgu@caep.cn, †weilu@ihep.ac.cn

**Abstract:** Wavelength-tunable ultra-intense femtosecond lasers may enable breakthroughs in diverse areas of science spanning attosecond science, particle acceleration and beyond. Conventional crystal-based methods are limited by gain bandwidth and damage thresholds, which restrict their wavelength tunability. Plasma-based frequency conversion, unconstrained by material damage, offers a promising alternative. Here, a novel scheme named Frequency Downshifting Stair (FDS) based on plasma bubble filling control is presented. The FDS enables arbitrary frequency down-conversion of ultra-intense femtosecond pulses and yields chirp-free laser pulses. It can achieve near-100% photon conversion efficiency, approaching the physical limit. This is attributed to the linear control by the FDS of laser chirp evolution during the photon deceleration in the plasma wake bubble. For a laser pulse with an arbitrary wavelength $\lambda_0$ (e.g., $\lambda_0 = 800$nm), proof-of-concept PIC simulations demonstrate that a single-stage FDS enables continuous wavelength tuning from $\lambda_0$ to $2\lambda_0$ (800–1600nm). Moreover, a three-stage cascaded FDS achieves more than tenfold frequency ($10\lambda_0$) downshifting to a central wavelength of 8.5µm. The FDS scheme thus provides a universal pathway for generating high-energy, few-cycle pulses across the broad infrared regime, offering a powerful new tool for wavelength-dependent ultrafast science.

## Introduction

Ultra-intense femtosecond lasers have catalyzed revolutionary advancements in diverse scientific domains including particle acceleration[1-3], ultrafast photochemistry[4-6], attosecond physics[7-14], among others. These frontier applications, however, place stringent demands on the wavelength tunability of the driving lasers further. In ultrafast photochemistry, wavelength-tunable lasers are essential for mode-selective excitation[15,16] and for steering ultrafast reaction dynamics[17]. In biomedical and surgical areas, wavelength-tunability also helps minimize collateral tissue damage by optimizing the interaction wavelength[11]. In high-harmonic generation (HHG), wavelength tunability offers a powerful handle to balance the trade-off between harmonic cutoff order ($\propto \lambda_L^2$) and conversion efficiency ($\propto \lambda_L^{-5 \sim -6}$)[18,19]. Beyond wavelength tunability, cycle-level duration and high pulse energy are also required in mid-infrared strong-field physics and single-shot ultrafast metrology[20]. These demands collectively highlight the need for high-energy, wavelength-tunable femtosecond lasers across the broad infrared regime.

Currently, nonlinear frequency conversion in optical crystals is the primary method to generate femtosecond laser pulses across the broad infrared band. It includes optical parametric amplification (OPA), optical parametric chirped-pulse amplification (OPCPA), and difference-frequency generation (DFG). However, the output laser wavelength and photon conversion efficiency ($\eta_p$) are tightly constrained by nonlinear crystals, with each spectral region typically requiring a dedicated laser system[21]. The $\eta_p$ of crystal-based techniques typically remains ~30% in the short-wavelength infrared (SWIR, 1.4–3μm) to mid-infrared (MIR, 3–8μm) regime[22-24]. In the SWIR regime, 2.4μm, ~150mJ pulses have been generated based on Cr:ZnS or BBO crystals. In the MIR regime, DC-OPA using MgO:LiNbO$_3$ has produced 3.3μm, 31mJ pulses, with $\eta_p \approx 22\%$ [25]. And beyond 5μm, limited crystal bandwidth necessitates more complex setups. For example, 0.65mJ pulses at 5.1μm are generated by a DFG-seeded OPCPA system using ZnGeP$_2$ crystals at 2μm pumping, though with $\eta_p = 6.1\%$[26]. $\eta_p$ would drop further as the target wavelength extends to long-wavelength infrared regime (8-15μm). The state-of-art pulses at 9μm (0.21mJ, 91fs) involve DFG driven by two sets of OPCPA systems operating at 2.3μm and 3.1μm, respectively, yet with only $\eta_p = 2.6\%$[27]. Obviously, the wavelength tunability constrained by optical materials, and the consequently low overall conversion efficiency, constitute the main limitations of this path.

Unlike solid-state crystals, plasma is free from bandwidth limitations and damage thresholds and can offer exceptional flexibility for manipulation. When an intense femtosecond laser pulse interacts with plasma, a plasma wake would be excited behind and co-propagates with the pulse. The spatial density distribution of each plasma oscillation period looks like a bubble where the laser displaces plasma electrons. Plasma wake bubbles have demonstrated significant potential for direct frequency down-conversion of high-peak-power femtosecond lasers known as photon deceleration[28-35]. For instance, Nie et al. use plasma wakes to first compress and then convert a sub-joule-level 800nm, 30fs laser. It generated tunable, near-single-cycle mid-infrared pulses spaning 5-14μm, achieving a $\eta_p$ of up to 25%[31,32]. Despite this, this plasma-based approach still faces two major limitations. First, the $\eta_p$ remains suboptimal and well below 100%. Second, the output laser generally exhibits complex spatiotemporal structures, such as nonlinear frequency chirps. Therefore, it's a central issue to achieve high-quality, wavelength-tunable laser generation from plasma with maximal $\eta_p$. A deeper understanding of photon deceleration physics may provide new insights.

In plasma wakes, the relative scale between the laser pulse and the plasma bubble is a key parameter in determining the frequency conversion performance. We introduce a bubble filling factor (BFF) to quantify how much of the plasma wake bubble for redshift is occupied by the laser pulse, defined as $BFF = c\tau_{laser}/R_b$, where $c\tau_{laser}$ is the laser pulse length and $R_b$ is the radius of the plasma bubble. Our study reveals a distinct effect transition in photon deceleration as the BFF varies. When $BFF \ll 1$(under-filling regime, e.g. $BFF = 0.2$), the plasma bubble acts primarily as a linear negative-dispersion medium for the laser pulse. As $BFF \approx 1$(fully-filling regime), the effect transitions to a quasi-linear positive-dispersion. The dual effects can be directly verifiable, both theoretically and via PIC simulations, since photon deceleration in plasma wakes can be described by a set of coupled nonlinear equations based on a cold-fluid model. These two BFF regimes are significantly distinct from the moderate regime ($BFF \approx 0.6$) typically employed in laser wakefield acceleration (LWFA)[36,37]. Building on this insight, is it possible to overcome the above limitations in photon deceleration by further integrating these two linear dispersion processes? We propose a practical frequency conversion scheme. In the scheme, a laser pulse first undergoes redshift and acquires a linear negative chirp within an under-filling regime. Subsequently, it enters a fully-filling regime and continues to redshift, introducing a compensating positive chirp which balances the initial dispersion. The down-converted laser pulse is expected to be chirp-free, so the scheme can be applied again for cascaded frequency down-shifting. This process is analogous to descending a stair, hence we term the scheme the "frequency down-shifting stair" (FDS).

In this work, we demonstrate the FDS scheme for wavelength-tuning of ultra-intense femtosecond lasers through particle-in-cell (PIC) simulations. The FDS enables complete and continuous wavelength-tuning from the near-infrared to long-wavelength infrared regimes with nearly 100% photon conversion efficiency. The required plasma-photonics structure is accessible, and has already been widely implemented in ultra-intense laser facilities. Owing to its plasma-based nature and absence of damage thresholds, the FDS offers a universal and efficient pathway toward high-pulse-energy, wavelength-tunable femtosecond laser sources.

## Basic concept

The modulation effect of the plasma wakes on the driving laser is mainly due to the asymmetric self-phase modulation[38]. As a laser pulse excites plasma wakes, its frequency is modulated by the gradient of refractive index $\partial \eta / \partial \xi$ in plasma. Additionally, plasma is a negative dispersion medium. Under the group velocity dispersion (GVD) of plasma, $v_g \approx c(1 - \omega_p^2/2\omega^2)$, long-wavelength components move slower. It is therefore called photon deceleration. The longitudinal non-uniformity and temporal evolution of plasma wakes cause the frequency modulation of the driving laser to vary along its propagation, making it difficult to achieve a uniform frequency redshift.

The FDS aims to solve the problem and achieve a uniform frequency shift through precise manipulation of the laser pulse's temporal-frequency evolution. The FDS process involves two distinct steps: the trailing-edge redshift step and the leading-edge redshift step. Fig. 1 illustrates the concept.

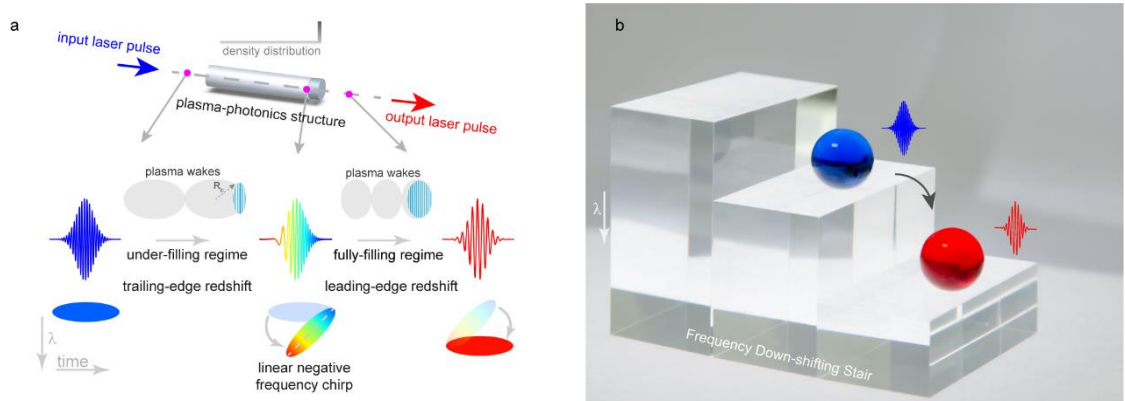

Fig. 1 Concept of the Frequency Down-shifting Stair (FDS). (a) the process of FDS. (b) Conceptual scheme. The purple dots in (a) mark the key transition points. The initial input laser pulse first undergoes a trailing-edge redshift in under-filling plasma bubbles to approach the target wavelength, followed by a leading-edge redshift in fully-filling plasma bubbles to match the target wavelength, $R_b$ is the radius of plasma bubble. The colorful laser pulse in the intermediate modulation state is a linear negative chirp pulse. This sequential process transforms the initial input laser pulse (blue pulse, λ₀) into a wavelength-tuned output laser pulse (red pulse, λ₁, where λ₁ > λ₀ and λ₁ is tunable), as shown in (b).

The trailing-edge and leading-edge steps are governed by plasma wakes in different filling regimes (see Methods). In Fig. 1a, during the trailing-edge redshift step, the under-filling plasma bubble modulation causes the laser's trailing edge to shift towards longer wavelengths, while the leading edge remains relatively unchanged. And the corresponding temporal-frequency distribution of the pulse exhibits a linear negative frequency chirp. Upon transitioning to the leading-edge redshift step, a fully-filling plasma bubble shifts the frequency of the leading part of the pulse towards a tunable target wavelength, $\lambda_1$. The resulting output pulse, as the red pulse in Fig. 1a, exhibits a chirp-free temporal-frequency distribution. In essence, a "blue ball" (initial short-wavelength laser pulse) can be directly converted into a "red ball" (final long-wavelength laser pulse) through FDS as a whole, as illustrated in Fig. 1b. Benefiting from the bandwidth-free nature of plasma, the initial and output wavelength in the FDS process is not specific and can be scaled. In the article, we take 800nm—a representative wavelength of femtosecond ultra-intense lasers—as an initial wavelength for demonstration.

## Results

**Demonstration of uniform frequency conversion**

First, we have demonstrated that FDS scheme enables continuous wavelength tuning of ultra-intense femtosecond laser pulses to arbitrary target wavelengths by performing PIC simulations. As shown in Fig. 2a, the laser spectrum undergoes a gradual redshift through the FDS process, culminating in a central wavelength shift to 1.3µm in the case.

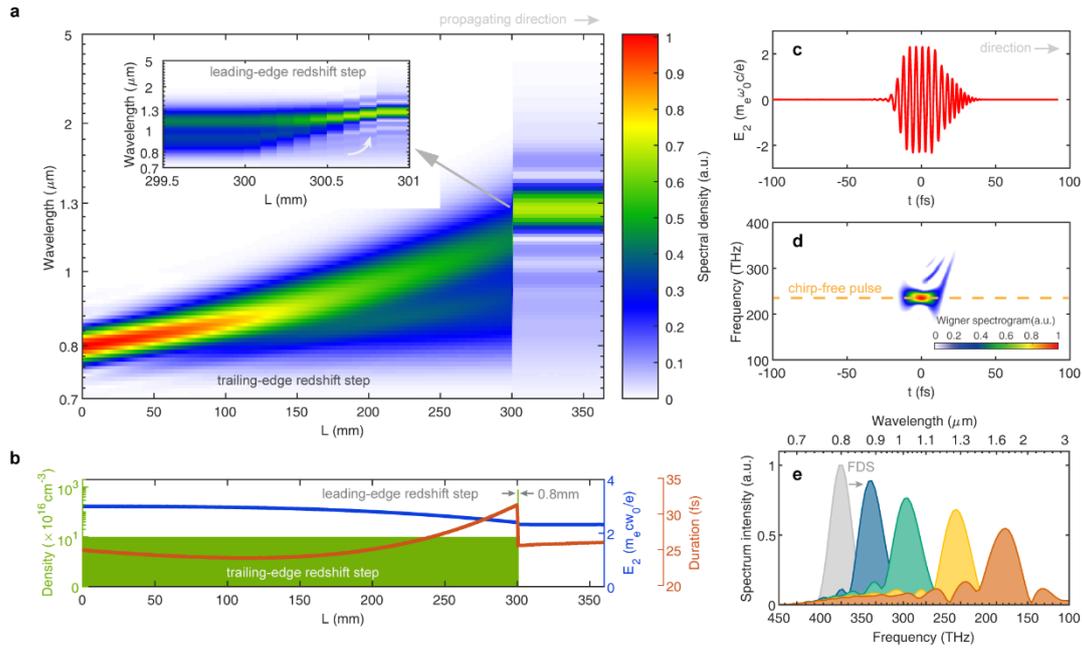

Fig 2. PIC simulation demonstration of FDS. (a) the spectrum evolution. The initial 800nm pulse is converted to 1.3μm. The insert is a local enlargement of the leading-edge redshift step; (b) the plasma-photonics structure and the evolutions of $a_0$ and the pulse duration; (c) $E_2$ field distribution of the output laser; (d) temporal-frequency distribution of the output chirp-free laser; (e) Adjusting plasma density or length enables continuous wavelength tuning. The spectral shift of an 800nm laser, spanning 0.8–1.6μm (0.9μm (blue),1.0μm (green),1.3μm (yellow) and 1.65μm (orange)) via FDS.

In detail, an 800nm input laser undergoes spectral broadening and redshifts in the trailing-edge redshift step. Then the portion of the spectrum that does not experience redshift before undergoes a rapid redshift in the subsequent leading-edge redshift step. Ultimately, the spectrum becomes concentrated around 1.3μm, achieving the desired overall frequency shift. From the perspective of spectral evolution, the laser spectrum is first broadened to simultaneously cover both the initial and the target wavelength, forming a supercontinuum spectrum. Then, the spectrum is narrowed towards the target wavelength, thereby achieving complete wavelength conversion. In Fig. 2b, the laser's peak-intensity slowly decreases, while the pulse duration experiences slight compression followed by significant elongation, reaching a maximum duration of 31fs by the end of the first step. In the leading-edge redshift step, rapid plasma modulation of the fully-filling bubble compresses the pulse back to 26fs. The output and input lasers exhibit similar durations and good quasi-monochromaticity. Fig. 2c shows the electric field distribution ($E_2$) of the output laser, confirming that the wavelength-tuned pulse still maintains high temporal quality. Fig. 2d presents the chirp-free temporal-frequency distribution of the output 1.3μm pulse.

The FDS scheme supports fine control over the output wavelength. By tuning the plasma density or length, the output wavelength can be continuously adjusted. Fig. 2e illustrates a spectral comparison, showcasing seamless tuning from $\lambda_0$ to $2\lambda_0$. Simultaneously, the spectral bandwidth of FDS-generated laser is notably broadened. For instance, at 1.3μm, the output laser exhibits a bandwidth (FWHM) of 196nm ($\Delta\nu = 36.5 \text{THz}$), a substantial increase compared to the initial 60nm bandwidth ($\Delta \nu = 27.8\text{THz}$) at 0.8μm. Combined with the chirp-free characteristic in the temporal-frequency distribution (Fig. 2d), the FDS enables an output pulse duration approaching the Fourier-transform limit. This spectral broadening is crucial for achieving near-single-cycle pulse durations.

**Stable temporal-frequency evolution of lasers in plasma wakes**

Temporal–frequency evolution is crucial for laser frequency modulation. Due to GVD in photon deceleration, the laser waveform will change as it propagates without intervention. The laser-driven plasma wakes are mainly determined by the laser ponderomotive potential, which is related to the pulse's waveform directly. Once the laser waveform changes, the distribution of the plasma wakes as well as the frequency modulation also vary. The lack of control over the temporal-frequency structure of the pulse, prevents the stable frequency-shifting via plasma wakes, which limits the conversion efficiency. This is a common challenge encountered across all existing photon deceleration approaches[31-34].

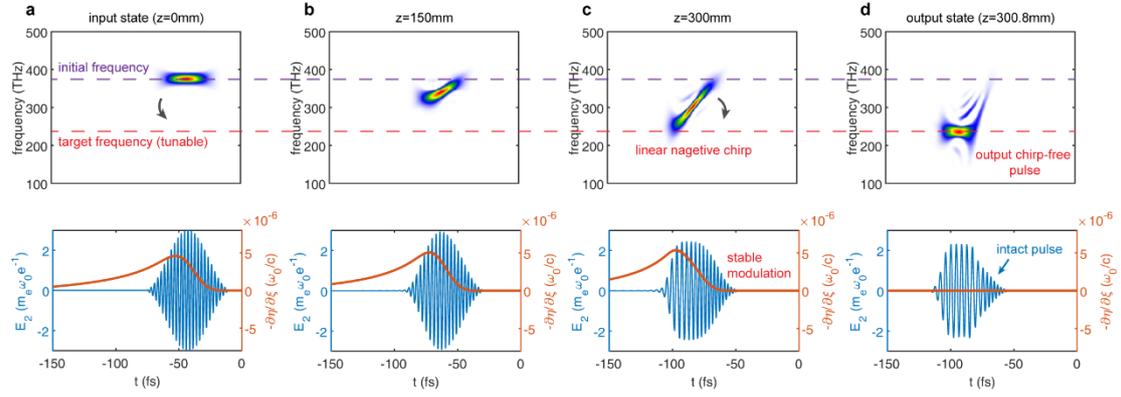

Fig. 3: (a-d) Evolution of laser waveform and temporal-frequency distribution in the plasma-photonics structure. In FDS process, the $\partial \eta / \partial \xi$ distribution relative to the laser profile remains nearly constant in each step, resulting in a smooth overall frequency shift without chirp, as shown in (d). The marked propagation z positions in the figure correspond to those shown in Fig. 2. The output target frequency supports tunable.

To ensure a stable temporal–frequency evolution of the pulse, the first step of FDS redshifts the trailing edge of the laser pulse, while leaving the leading half nearly unchanged. As a result, the plasma wakefield, driven by the ponderomotive force of the laser, remains stable during propagation. This enables the laser pulse to undergo steady redshift over extended plasma distances. Once the trailing portion reaches the target frequency, the second step of FDS rapidly redshifts the leading edge to the desired wavelength.

The stable evolution is clearly visualized in Fig. 3. The $\partial \eta / \partial \xi$ is stable and remains unchanged over the whole time respectively in either redshift step. First, during the trailing-edge redshift step (Fig. 3a-c), under the stable $\partial \eta / \partial \xi$ modulation, the laser's trailing-edge stably redshifts, forming a linear negative chirp. Meanwhile the frequency of the leading-edge remains unchanged. Then in the leading-edge redshift step (Fig. 3c-d), the rapidly redshifts the laser's leading part, and forms a chirp-free pulse with wavelength at 1.3μm in Fig. 3d eventually. Comparing Fig. 3d with Fig. 3a, the driving laser has been transformed from an initial chirp-free pulse, to a chirp-free state as well with a new wavelength. From the perspective of laser's temporal-frequency distribution, the FDS effectively enables the complete translation of a short-wavelength laser pulse to a longer wavelength regime, as exemplified by the transition from the purple line to the red line in Fig. 3.

The output wavelength supports tunability. The stepwise approach, employed by FDS, allows for stable and precise control over both the frequency shift $\Delta f$ and frequency chirp, facilitating the efficient generation of chirp-free laser pulses at desired frequencies. Specifically, the lack of control over $\Delta f$ can lead to longitudinal fragmentation of the broadened pulse due to GVD, disrupting the wake frequency modulation and further constraining efficiency and tunability. To limit GVD, empirically, controlling $\Delta f \lesssim f_0/2$ can yield optimal results in a single FDS stage, where $f_0$ is the initial laser frequency.

**Cascaded FDS to generate wavelength-tunable intense single-cycle MIR/LWIR lasers**

The FDS supports multi-stage cascading to extend the output laser wavelength further into MIR and LWIR regime. In one FDS stage, the output chirp-free laser maintains high quality, as detailed in Fig. 3d, which is similar to the input laser state and is capable to excite plasma wakes again. Thus, mid-infrared lasers could be achieved naturally by subjecting the laser pulse output from one FDS process to another FDS round. Regarding the output wavelength, multi-stage cascaded FDS could even output LWIR laser pulses. Fig. 4 illustrates the PIC simulation result of cascaded FDS and the parameters are summarized in Tab. 1. Since each FDS stage includes both a trailing-edge redshift step dominated by under-filling plasma bubble and a leading-edge redshift step by fully-filling plasma bubble, the plasma layout of cascading FDS seems to form a periodic density structure as the gray columns shown in Fig. 4a.

The prominent feature of FDS is its stair-like overall frequency tunability of femtosecond lasers. According to the three-stage cascaded FDS simulation, an initial $\lambda_0 = 800\mathrm{nm}$ laser pulse is sequentially shifted to $2\lambda_0$, $4.5\lambda_0$, and $10.6\lambda_0$ in Fig. 4b. The spectrum after each FDS stage exhibits quasi-monochromatic characteristics, resembling a translation of the peak spectral wavelength. The cascading frequency conversion in Fig. 4b demonstrates the "frequency stair" concept depicted in Fig.1a. Besides, leveraging the continuous wavelength tunability of single-stage FDS verified in Fig. 2e, multi-stage FDS can achieve fully tunable output spanning the NIR to LWIR spectrum.

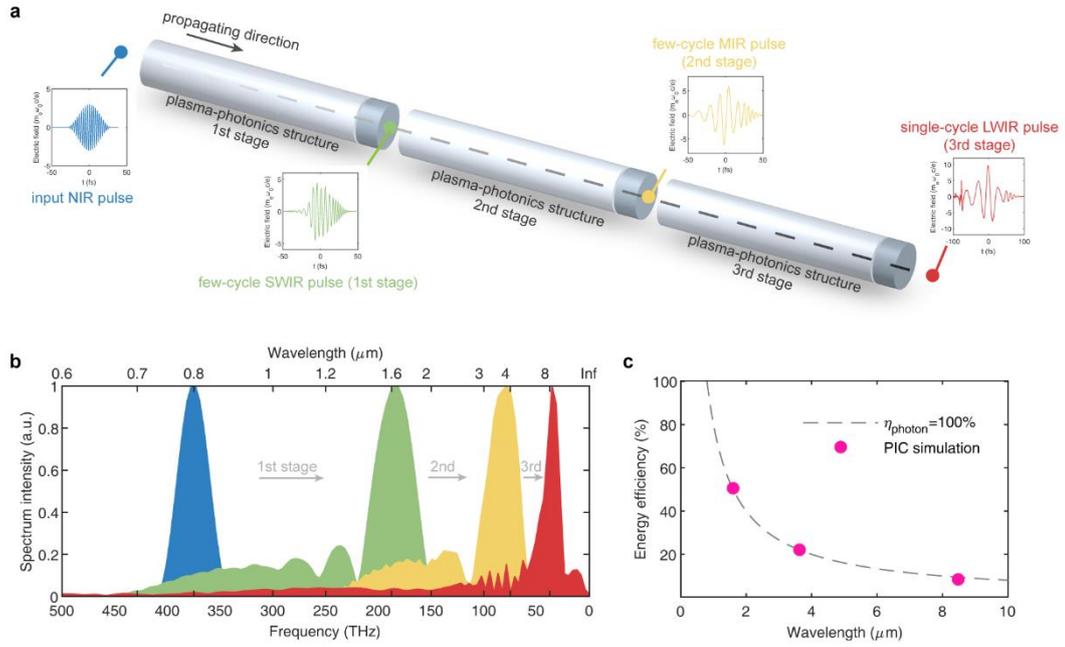

Fig. 4: PIC simulation of cascaded FDS to generate tunable single-cycle mid-infrared lasers. (a) Plasma configuration for the cascaded FDS process. The inset shows $E_2$ field distribution of lasers from each stage. (b) Output spectrum of lasers at 1.6μm, 3.6μm, and 8.5μm from each FDS stage. (c) the energy conversion efficiency from PIC simulations based on an initial 800nm laser pulse. As the pulse progresses through the plasma-photonics structure, its duration approaches a single cycle, with three-stage FDS cascading supporting wavelength output in the LWIR region.

Such quasi-monochromatic conversion feature enhances the laser conversion efficiency significantly. Fig. 4c shows the total cascading energy conversion efficiencies (the purple dots) from PIC simulations. The dashed gray line represents 100% photon conversion efficiency, indicating the theoretical maximum efficiency where all initial laser photons are converted to the target frequency. It can be observed that the simulated energy conversion efficiency closely matches the theoretical maximum. Specifically, Using FDS scheme, approximately 50% of the energy from an initial 800nm femtosecond laser pulse is converted to a 1.6μm femtosecond laser. In contrast, conventional optical pump-amplification techniques, using 800nm Ti:sapphire lasers as the pump source, generate 1.7μm femtosecond lasers, achieving typical energy conversion efficiencies of 14%[24,39]. This highlights a significant advantage of FDS over traditional crystal pump-amplification in terms of energy conversion efficiency. Additionally, the energy conversion efficiency of the cascaded FDS scheme is noteworthy. As illustrated in Fig. 4, the energy efficiency for converting the femtosecond laser from 800nm to 3.6μm is 22%, while the efficiency from 800nm to 8.5μm remains at an 8.4%. It means single-cycle LWIR lasers centered at 8.5μm with ~1J pulse energy and tens of terawatt peak-power can be produced from a 10J, 800nm femtosecond laser.

The FDS output pulses maintain a pulse duration similar to that of the input pulse. As the output wavelength increases, the number of optical cycles within the pulse gradually decreases, resulting in few-cycle SWIR and MIR pulses. When the output wavelength reaches 8.5μm, it even generates single-cycle LWIR laser pulses as indicated in Fig. 4a.

Tab. 1 PIC simulation parameters of cascaded FDS process

| stage | Input laser | | Trailing-edge | | Leading-edge | | Output laser | |
|---|---|---|---|---|---|---|---|---|
| | $\lambda$ (μm) | $a_0$ | $n_{e1}$ (cm$^{-3}$) | $L_1$ (mm) | $n_{e2}$ (cm$^{-3}$) | $L_2$ (mm) | $\lambda$ (μm) | $a_0$ |
| 1st | **0.8** | 3 | $2 \times 10^{17}$ | 125 | $8 \times 10^{18}$ | 1.0 | **1.6** | 4.5 |
| 2nd | **1.6** | 4.5 | $8 \times 10^{16}$ | 90 | $4 \times 10^{18}$ | 0.6 | **3.6** | 6.4 |
| 3rd | **3.6** | 6.4 | $2 \times 10^{16}$ | 100 | $4 \times 10^{18}$ | 0.2 | **8.5** | 9.7 |

The final highlight of FDS is the enhancement of the normalized laser intensity $a_0$, given by $a_0 = 0.85(I\lambda_{\mu m}^2/10^{18}\text{Wcm}^{-2})^{1/2}$, where $I$ is the laser intensity and $\lambda$ is the laser wavelength. A large $a_0$ offers significant benefits for relativistic effects. Through FDS, the increased output wavelength leads to an enhancement of $a_0$. As detailed in Tab. 1, starting from an $a_0 = 3$, 800nm pulse, the three-stage FDS produces a single-cycle 8.5μm laser pulse with a three times higher $a_0$. The scaling law for $a_0$ of FDS output pulses is $a_0' \propto \lambda_{out}^{1/2}$, where $\lambda_{out}$ is the output laser wavelength. This scaling law aligns well with PIC simulations. Those characteristics of FDS including wavelength-tunability, single-cycle duration and $a_0$-boosting, are attractive from application perspectives.

## Discussion

Until now, we present a novel approach, the FDS, for achieving arbitrary wavelength transformation of ultra-intense femtosecond lasers with nearly 100% photon conversion efficiency. This spectral broadening-narrowing method enables continuous tunability of the output central wavelength in the broad infrared regime. We identify two opposing linear dispersion modes induced by plasma wakes during the modulation of the driving pulse: one generates a negative frequency chirp, while the other produces a positive chirp. Through a plasma-photonics structure combining these two modes, FDS produces chirp-free laser pulses, with nearly all photons concentrated at the wavelength of interest. The near-100% photon conversion efficiency of FDS greatly enhances its potential for applications. The feasibility of the approach has been confirmed through PIC simulations, demonstrating the successful wavelength shifting of femtosecond lasers from 800nm to SWIR, MIR and LWIR regions.

It is worth noting that the FDS scheme is already feasible given the current state of laser and plasma technologies. First, femtosecond lasers with peak-power ranging from terawatt to petawatt have been widely established and even commercialized globally. As for the plasma structure, in trailing-edge step, a parabolic plasma channel might be necessary to effectively guide the laser[40]. The plasma length could be shortened to few centimeters by increasing the plasma density properly. The plasma in the leading-edge step does not introduce any additional challenges. Further, we point out that the plasma could be a continuous step-like whole ensemble or two separated uniform plateau-like plasma segments. The good news is the plasma-accelerator community has already developed such plasma structures[32,41-43]. Current research has demonstrated the experimental generation of plasma channel structures extending several meters, achieved through methods such as discharge capillaries or Bessel-beam ionization. These advancements significantly surpass the plasma conditions required for FDS implementation.

Looking ahead, the FDS scheme offers a universal method for tuning the wavelength of ultra-intense femtosecond lasers. FDS can transform single-wavelength "monochromatic" ultra-intense femtosecond lasers into wavelength-tunable "polychromatic" lasers, significantly broadening their application potential. Since the FDS scheme imposes no restrictions on the initial laser wavelength, it offers substantial potential for future advancements. As broadband, high peak-power ultrafast laser technologies mature across new spectral ranges, such as Thulium-doped femtosecond lasers operating at 2μm, FDS could enable the generation of cycle-level ultra-intense lasers spanning visible, mid-infrared, long-wave infrared, and even terahertz regimes more directly. We envision that the FDS scheme could make wavelength tuning as straightforward as adjusting pulse energy. As a plasma-based approach, the FDS is adaptable to ultrafast laser facilities of varying power levels, meeting the wavelength-tuning needs of a wide range of femtosecond laser applications.

## Methods

**Theoretical analysis of laser frequency modulation in the plasma-photonics structure.** In FDS scheme, we discovered that the laser pulse dispersion effect with linear chirp can be achieved by plasma bubble filling control. Photon deceleration can be described by a set of coupled nonlinear equations based on a cold-fluid model with a quasi-statistic approximation. The theoretical form of the one-dimensional plasma wake is governed by Equation (1) and (2)[44]:

$$\left(\frac{2}{c}\frac{\partial}{\partial \xi} - \frac{1}{c^2}\frac{\partial}{\tau}\right)\frac{\partial \boldsymbol{a}(\xi)}{\partial \tau} = k_p^2 \frac{\boldsymbol{a}(\xi)}{1+\phi} \qquad (1)$$

$$\frac{\partial^2 \phi}{\partial \xi^2} = \frac{k_p^2}{2}\left[\frac{1+a^2}{(1+\phi)^2} - 1\right] \qquad (2)$$

Here, $\xi = z - ct$, $\tau = t$, $\phi = |e|\Phi/m_e c^2$ is the normalized scalar potential, and $\boldsymbol{a}$ (defined as $a_0 = eE_L/m_e c\omega_L$) represents the laser normalized vector potential. For a given $a(\xi)$ and plasma parameter $k_p^2$, Equation (2) allows the numerical computation of the normalized scalar potential distribution $\phi(\xi)$. Once $\phi(\xi)$ is determined, the electron density distribution $n(\xi)$ in the wake could be known. Consequently, the refractive index distribution $\eta(\xi) = [1 - n_e(\xi)/n_c]^{1/2}$ can be calculated, where $n_c = m_e \omega^2/(1+\phi(\xi))$ and the relationship between refractive index distribution $\eta(\xi)$ and $\phi(\xi)$ is given by:

$$\eta(\xi) = 1 - \frac{\lambda^2/2\lambda_p^2}{1+\phi(\xi)}, \qquad (3)$$

Assuming a stable propagation of the laser in the plasma over a certain distance $L$, the instantaneous phase of the laser $\varphi(t) = \omega_0 t - kz = \omega_0 t - 2\pi\eta(\xi)L/\lambda_0$. The refractive index gradient $\partial\eta/\partial t$ induces phase modulation on the laser. Under the quasi-static approximation, $\partial\eta/\partial t$ is approximately given by:

$$\frac{\partial\eta}{\partial t} \simeq -c\frac{\partial\eta}{\partial\xi} = -c\frac{\lambda^2/\lambda_p^2}{1+\phi(\xi)}\frac{\partial\phi(\xi)}{\partial\xi}, \tag{4}$$

The frequency of the laser pulse $\omega(t)$ after interaction with the plasma can be expressed as:

$$\omega(t,\xi) = \frac{d\varphi(t)}{dt} = \omega_0 + \frac{2\pi c L \lambda_0}{\lambda_p^2}\frac{1}{[1+\phi(\xi)]^2}\frac{\partial\phi(\xi)}{\partial\xi}, \tag{5}$$

According to Equation (5), the frequency shift $\Delta\omega$ of the laser pulse is given by:

$$\Delta\omega = \omega(t) - \omega_0 = \frac{2\pi c L \lambda_0}{\lambda_p^2}\frac{1}{[1+\phi(\xi)]^2}\frac{\partial\phi(\xi)}{\partial\xi}, \tag{6}$$

The direction of the frequency shift, i.e., whether the laser pulse undergoes redshift or blueshift, depends on the sign of $\partial\phi(\xi)/\partial\xi$. When $\partial\phi(\xi)/\partial\xi < 0$, $\Delta\omega < 0$, indicating a decrease in the laser pulse frequency, corresponding to photon deceleration. When $\partial\phi(\xi)/\partial\xi > 0$, $\Delta\omega > 0$, indicating an increase in the laser pulse frequency, corresponding to photon acceleration. Since the modulation is stable in FDS scheme, the frequency shift of the output laser relative to the initial pulse can be roughly calculated using Equation (6).

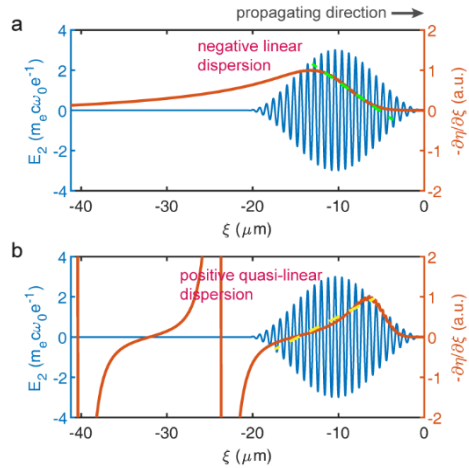

Fig. 5: Theoretical solution of modulation effects on laser pulse ($a_0 = 3$, 25fs) in plasma wakes. (a) Under-filling regime in trailing-edge redshift step ($n_e = 1 \times 10^{17} \text{cm}^{-3}$), where the plasma wake predominantly induces redshift and a negative linear dispersion. (b) Fully-filling regime in leading-edge redshift step ($n_e = 8 \times 10^{18} \text{cm}^{-3}$), where the plasma wake primarily introduces a quasi-positive linear dispersion on the main body of the pulse.

The frequency shift $\Delta\omega$ at different positions $\xi$ of the laser pulse is associated with the distribution of $\phi(\xi)$ in Equation (6). In essence, by manipulating plasma wakes, we find diverse effects on laser frequency shift can be achieved. Fig. 5 illustrates two operational modes of the plasma wakes on the driving laser, numerically calculated via the theoretical model of one-dimensional plasma wake outlined in Equation (2). Firstly, Fig. 5a portrays the trailing-edge redshift mode in an under-filling plasma bubble. In this mode, the plasma wake induces negative linear dispersion on the driving laser, resulting in a redshift in the laser pulse's trailing-edge while maintaining the frequency of the leading-edge relatively invariant. An initially chirp-free laser pulse, after undergoing the trailing-edge redshift mode, would transform into a negatively chirped pulse with a redshifted spectrum. In contrast to that, Fig. 5b outlines the leading-edge redshift mode in a fully-filling plasma bubble. In this configuration, the plasma wake induces leading-edge redshift and an approximately positive linear dispersion. Thus, such two redshift modes form the basis of FDS concept as depicted in Fig. 1. The trailing-edge redshift mode maintains its role in inducing a linear negative chirp dispersion, causing a redshift in the initially chirp-free laser pulse. The key distinction lies in the subsequent action of the leading-edge redshift mode, which introduces a quasi-positive linear chirp dispersion to the laser pulse. This transformation renders the pulse chirp-free, achieving an integrated frequency shift in the laser pulse.

**PIC Simulations.** The PIC simulations were carried by the OSIRIS code. As for the parameters of the demonstration case in Fig.2, the input laser has a central wavelength of 800nm, normalized intensity $a_0 = 3$, and a pulse duration of 25fs (Full Width at Half Maximum, FWHM). In terms of the plasma-photonics structure, the plasma density is $n_e$=1×10$^{17}$cm$^{-3}$ with a length of 30cm in the trailing-edge redshift step. In the leading-edge redshift step, $n_e$ =8×10$^{18}$cm$^{-3}$ and the length is 0.8mm as detailed in Fig. 2b.

## Acknowledgements

This work was supported by the Strategic Priority Research Program of the Chinese Academy of Sciences (Grant No. XDB0530000), the National Natural Science Foundation of China (Grants No.12235014 and No. 12405287), the Foundation of National Key Laboratory of Plasma Physics (Grant No. 6142A04240101), the Discipline Construction Foundation of "Double World-class Project". The simulation work is supported by the Center of High Performance Computing, Tsinghua University.

## Author contributions

Y.H. and W.L. proposed the concept. Y.H. developed the theoretical model and carried out the simulations. Y.H., X. N., B. G., J.H. and W.L. wrote the paper. W.L., J.H., Y.G. conceived and supervised the project. All authors discussed extensively the results and commented on the manuscript.

## Competing interests

The authors declare no competing interests.